# DETECTING CONCRETE ABNORMALITY USING TIME-SERIES THERMAL IMAGING AND SUPERVISED LEARNING

**Chongsheng Cheng**
The Durham School of Architectural Engineering and Construction
University of Nebraska-Lincoln
113 Nebraska Hall, Lincoln, NE 68588-0500
E-mail: cheng. chongsheng@huskers.unl.edu

**Zhigang Shen, Ph.D., Corresponding Author**
The Durham School of Architectural Engineering and Construction
University of Nebraska-Lincoln
113 Nebraska Hall, Lincoln, NE 68588-0500
E-mail: shen@unl.edu

Word count: 4,132 words text + 10 tables/figures × 250 words (each) = 6,632 words

**ABSTRACT**

Nondestructive detecting defects (NDD) in concrete structures have been explored for decades. Although limited successes were reported, major limitations still exist. The major limitations are the high noises to signal ratio created from the environmental factors, such as cloud, shadow, water, surface texture etc. and the decision making still relies on the engineering judgment of interpretation of image content. Time-series approach, such as principle component thermography approach has been experimented with some improved results. Recent progress in image processing using machine learning approach made it possible for detecting defects' thermal features in more quantitative ways. In this paper, we provide a procedure to represent the thermal feature in the time domain by principal component analysis and regress the prediction of detection by two schemes of supervised learning models. Three independent experiments were conducted in a similar laboratory setup but varied in conditions to illustrate the performance and generalization of models. Results showed the effectiveness for the detection purpose with appropriate tuning for parameters. Future studies will focus on implementing more sophisticated structured models to handle more realistic cases under natural conditions.

## INTRODUCTION

Detecting sub-surface defects (e.g. delamination and void) of a concrete structure (e.g. bridge deck) plays a vital role in keeping structural integrity. Thermography as one of the non-destructive detection methods for concrete structures has been developed for decades. It features the potentials of high efficiency for large area scanning and no interruption to the traffic comparing to the conventional hammer sounding and chain dragging procedures. However, thermography as a product from optical imaging system suffers the degraded performance from environmental noise (*1; 2*) and the complicate interpretation of image content. As a result, the engineering judgment is always required to determine the defected areas based on the interpretation of data. Recent studies had been conducted for improving the performance of detection including the reliability analysis to trade off threshold selection by Sultan and Washer (*3*) and k-means clustering by Omar and Nehdi (*4*); Omar, Nehdi and Zayed (*5*). These proposed methodologies provided solutions at some degrees automatic procedure for delamination discrimination, but the judgment was still needed for deciding the trade-off level or selecting the number of k means and no easier decision comparing before. This paper tries to answer the question that can we find a generic way to achieve automatic detection for the defects of concrete structure captured by thermal imaging system? In a more specific way, can the feature be distinguishable for suspect areas (debond) from solid areas (health) by using a structured model representation? This problem could be treated as the classification problem from the point view of machine learning and able to be handled by utilizing supervised learning procedure.

The practical implementation of thermography to detect delamination-liked defects is generally based on the contrast method. The principle of this method relies on the phenomenon of the spatial surface temperature difference between solid and debonded areas across a daily thermal cycle. According to Hiasa et al. (*6*), concrete structure (e.g. bridge deck) on the field would experience two phases: heating phase and cooling phase. During the heating phase, the debonded area was heated by the solar radiation faster than solid area and thus presented as a hotter region; while during the cooling phase, it cooled down faster and presented as a cooler region than surroundings. The contrast method came into the way to threshold the debonded area based on this temperature difference (*7*). Thus, at the right time of collecting the data and selecting the appropriate cut-off level became the essence of this method (*2; 3; 7; 8*). More importantly, the behavioral difference of temperature between the debonded and solid areas did exist across time and was not utilized for distinguishing purpose yet due to the nature of contrast-based method which only focuses on the single thermal image. Thus, the purpose of this paper is further refined to investigate the time-series features of the transient heat behavior by introducing two schemes of supervised learning implementations in a case study for void detection of the concrete block under the laboratory condition.

## BACKGROUND

### Selecting Time-Series Thermography as The Feature Representation

In the field of infrared physics, there were several time-series based methodologies have been widely used to improve the detection performance by representing the spatial feature of thermal property from the time domain. Methodologies such as Fourier transform based Pulse Phase Thermography (PPT) and Principle Component Thermography (PCT) have been reported efficient in terms of the inhomogeneous sample surface and non-uniform heating under experimental investigation (*9*). But there were limited studies conducted to apply them for the on-field application. Arndt (*9*) adapted the square pulse phase thermography (PPT) for detecting the delamination-like defects inside the concrete slab for qualitative and quantitative investigation. The depth information for defect could be revealed by the phase image in the frequency domain. Van Leeuwen, Nahant and Paez (*10*) studied the PPT for honeycombing defects detection in concrete structures both under artificial heating and solar heating conditions. Dumoulin, Crinière and Averty (*11*) developed a thermal imaging system for bridge inner structure condition monitoring. The system revealed the inner structures (e.g. beam and girders) layouts underneath the deck by using PPT and PCT analysis. Ibarra-Castanedo et al. (*12*) extended the applicability of PPT and PCT for detecting hidden structure inside exterior wall under long-term natural solar loading as the excitation source. It also provided a theoretical support based on thermal wave theory. The essential phenomenon was the observed phase delay of temperature evolution occurred for different inner structures during the thermal cycle for days. In (*13*), the performance of time-series method had been evaluated and compared for detecting the voids in concrete blocks during heating stage under the experimental setup. From above literatures and experimental study in (*13*), we could observe several facts: (1) temperature variate differently for debonded and solid areas both in the indoor and outdoor environment; (2) this feature of variation across time could be utilized as a distinguishing feature when time-series methods were adopted. Thus, it makes sense that instead of focusing on the temperature contrast of a single thermal image to distinct debonded and solid areas at a moment of time, we could represent them by the feature of temperature variation across time.

**Machine Learning on Time-Series Thermography for Sub-Surface Defect Detection**

Limited numbers of studies were found to apply machine learning technologies for the purpose of sub-surface defect detection. At an early stage, Maldague, Largouët and Couturier (*14*) implemented perceptron network for defect detection based on PPT for an aluminum plate. In the architecture, 32 neurons were used as input layer to account 32 phase values for each pixel and a single neuron for the output layer. Later, two separate neural networks were developed for detection and depth estimation for composite material (e.g. carbon fiber reinforced plastic) by Darabi and Maldague (*15*) under the active thermography process. The input data used was either raw temperature or post-processed values (e.g. temperature contrast, PPT based phase value). The architectures of network remained on multilayer perceptron networks. Recently, more detailed studies on neural network architectures were conducted by Dudzik (*16*) and Halloua et al. (*17*). The classification neural network was found suitable for detection and the regression neural network was suitable for depth estimation. Kaya and Leloglu (*18*) implement several supervised classifiers to detect different types of buried landmines based on the feature of surface temperature evolution across a day. In so far, there is no such study conducted for the delamination/void like defect within the concrete structure under the circumstance of natural solar excitation. However, this circumstance is the key to field inspection. Before direct implementation on a real word case under the uncontrolled environmental condition, an experimental study is desired to investigate the applicability of methodology with a simple sample under a controlled environment.

**EXPERIMENT AND DATA PREPARATION**

The experiment was conducted to collect sequential thermal images for the hollow cored concrete masonry unit (CMU) under the artificial heating condition. Figure 1 showed test samples (a), layout (b), and the thermal image took at end of the heating stage (c). The sample blocks were heated by halogen lamp with 600-Watt power and data was recorded by the infrared camera (FLIR

A8300). The infrared camera has a maximum spatial solution of 1280×720 pixels with the thermal sensitivity of 20 mK.

Collected thermal images were raw data of temperature values and several prepossessing were conducted. Firstly, the data was stored in a 3D matrix that reserved the first two dimension for spatial locations and the third dimension for the time sequences. Then a spatial reduction was conducted to average out the random noise and decrease the requirement for computational power. Thus, the original size of the data was reduced from 360×640×360 to 90×160×360 when using the 4×4 pixel window. Under this process, the size of each pixel was equivalent to 40 (cm)/138 (pixels) = 0.29 (cm per pixel horizontal) which was still sufficient for the spatial resolution. Finally, three data sets were selected from three independent experiments under similar environmental condition. All data sets were collected for heating phase from room temperature and differed in the starting temperature and duration. One dataset was used for training and the other two were used for further validation. More detail was discussed in the later section.

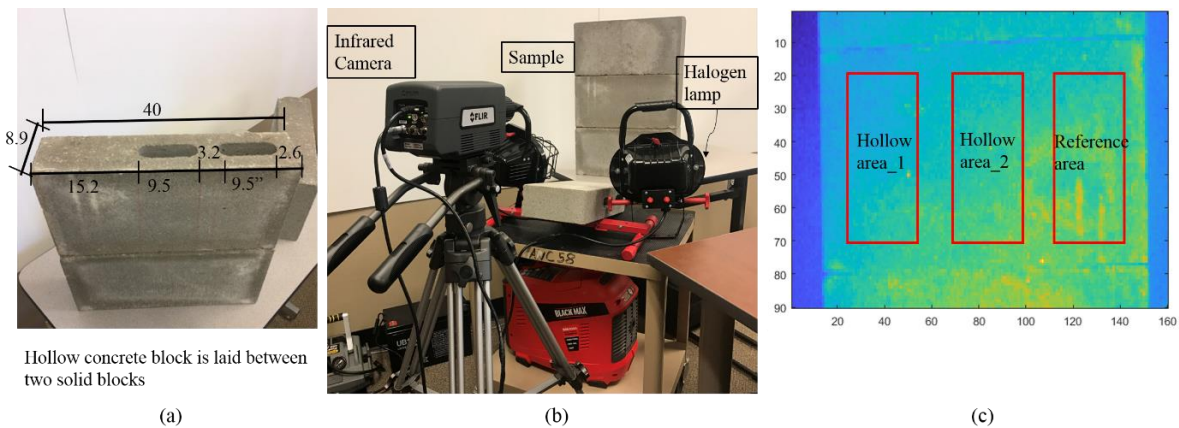

**Figure 1. Experiment setup: (a) sample dimension (cm); (b) layout; (c) raw image at end of heating**

## FEATURE DIMENSION REDUCTION AND LABELING

Once data was prepared, the principal component analysis (PCA) was used for feature dimension reduction. First, the data structure was reformatted into a 2d roaster matrix which column represented the thermal image as the vector form and row represented frames. For this case, we had a matrix with the size of 360×14400 which represented 360 time-series features and 14400 (90×160) pixel samples. Accordingly, the question could be reformed as that with these 360 time-series features, could we categorize 14400 pixels into two groups (solid areas and areas with hollowed-core underneath)? Figure 2 illustrated the dimension reduction by PCA and labeling process. PCA was conducted on the roaster matrix to reduce the feature dimension and was calculated by the singular vector decomposition (SVD). By doing so, the feature of temperature evolution in time domain was projected to the variation axes so that the most variation from data could be stored in fewer orthogonal dimensions with keeping minimum errors. This process was the same procedure for the method of principal component thermography (PCT) in the literature (19). Labeling process was used to pre-defined the group of pixels. Here we assign 1 to the pixels belonging the areas with voids and 0 to the pixels belongs to the solid areas. Once the feature deduction and labeling were determined, the decision of selecting the number of new features became the next step. Based on authors' experiment and observation, the first ten principal components were selected based on (1) more than 95% percent variation was explained by first ten components; and (2) random noise more intended to occur at the later principle components.

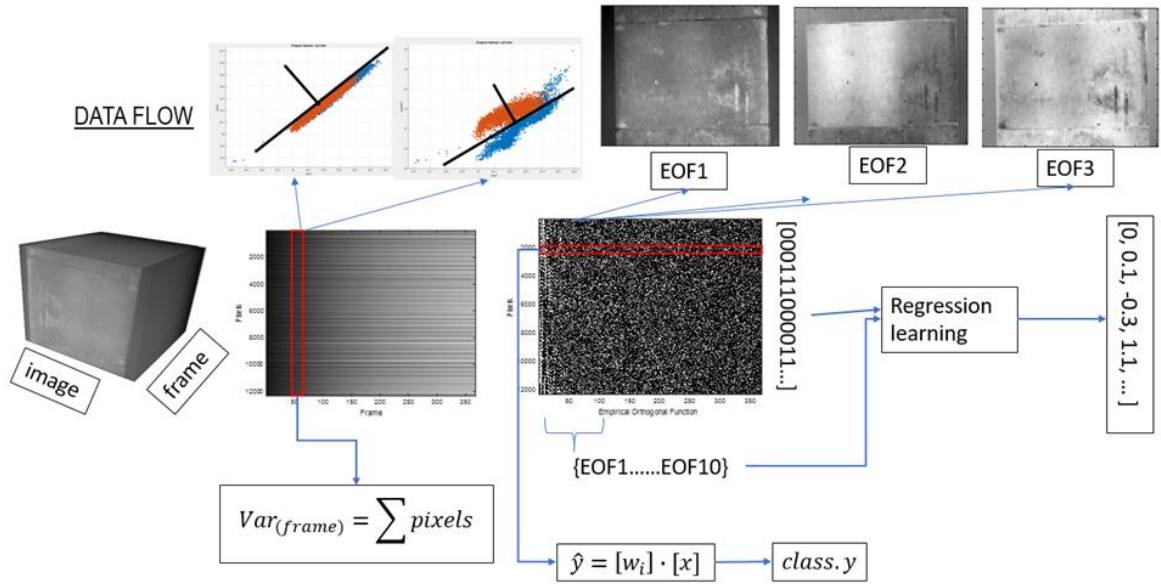

**Figure 2. Dimension reduction by PCA and labeling process**

## MODEL SELECTION AND DEVELOPMENT
### Supervised Training and Prediction Procedure

Figure 3 illustrates the procedure for model training, prediction, and validation. Firstly, the training dataset was selected, and the regions were determined for labeling. For our case, two yellow regions were labeled as 1 to indicate there were voids underneath and three blue regions were marked as 0 for solid areas. Then, the PCA was followed in Figure 2 so that we would have the feature matrix with labels to fit the learning models. Two schemes of regression models were used here to achieve the detection purpose: regression learner and regression neural network. The regression process selected here was based on the ideas that instead of using classifier for rigidly categorizing samples into binary or multiple classes, we could regress the predictors (features) to the response (labels) as the boundary from 0 to 1. As the result, we could have a softer outcome for the "classification" at this stage. Finally, with trained models, another two independent datasets were used for further evaluations.

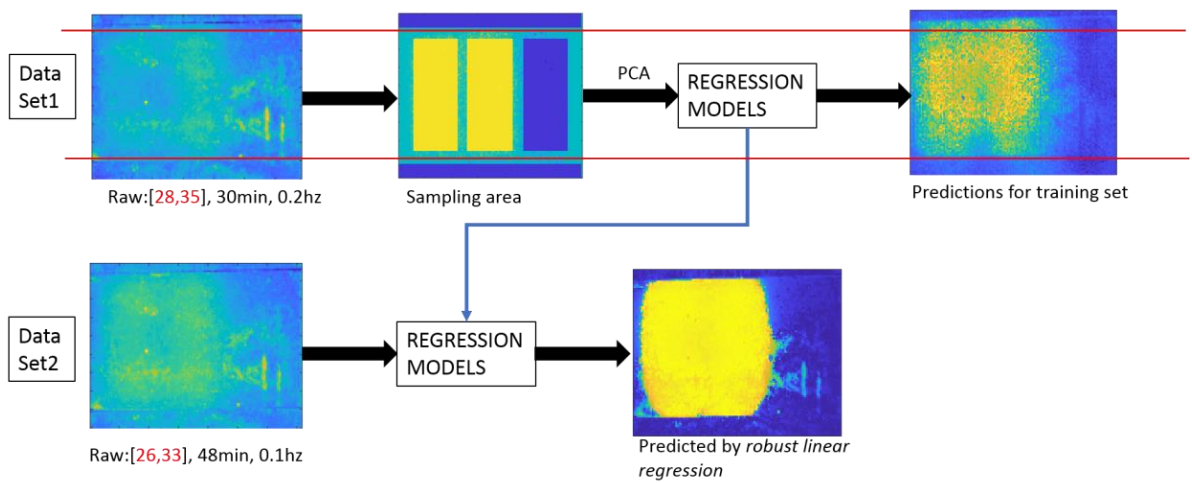

**Figure 3. Data flow for training and prediction procedure**

Three datasets were collected from three independent experiments to test the robustness of selected models. All experiments were used the same sample blocks and conducted in the laboratory environment with similar setups but differed in the duration of heating, the power of the heating source, and variation of room temperature. Figure 4 (a) showed the average temperature evolutions for samples across frames (time). Dataset 1 (Figure 4. b) was recorded at 0.2Hz sampling rate and the average temperature was from 28 ℃ to 35 ℃. Total of 360 frames was used and summed up 30 minutes for the heating process. Dataset 2 (Figure 4. c) was recorded at 0.1 Hz sampling rate from 26 ℃ to 33 ℃ with 292 frames used (total 48 minutes for heating). Dataset 3 (Figure 4. d) was recorded at 0.2 Hz sampling rate for 200 frames (16 minutes) from the temperature of 25 ℃ to 27 ℃. The additional difference of dataset 3 was the position of hollow-cored flipped opposite to dataset 1 and 2. Recall that only dataset 1 was used for training the models and dataset 2 and 3 were directly fitted into the trained model. Thus, the outcome of prediction from dataset 2 and 3 could reveal the robustness of the detection in terms of accounting the different temperature variations behaviors.

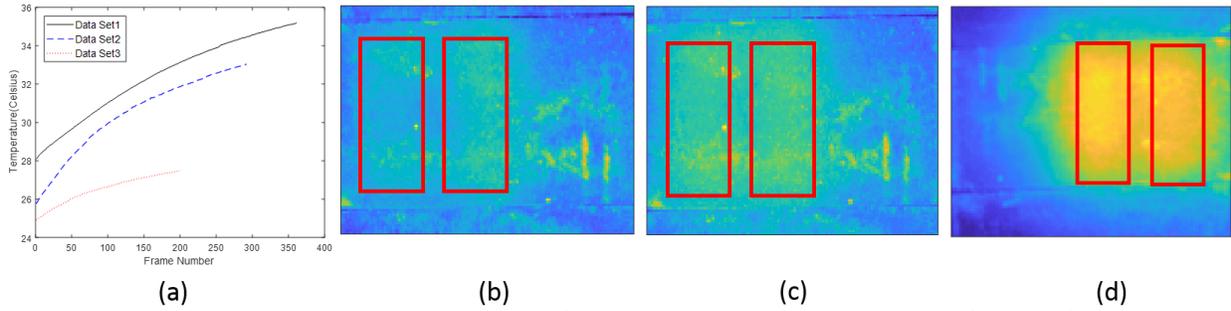

**Figure 4. (a) Average temperature evolution for 3 datasets; raw image at last frame of each dataset: (b) dataset 1; (c) dataset 2; and (d) dataset 3. Red boxes indicated the areas with hollowed cores underneath.**

### Regression Learner Models

Multiple regression learners were tested in the study including linear regression, support vector machine (SVM), and decision tree. A MATLAB built-in package was used to conduct the training and evaluation (ten folds cross evaluation) process. A standard scaling process was conducted to normalize the features after the PCA process. The intuition of regression model could be considered as follows: suppose existing a model $H_\theta$ that made $\hat{Y} = H_\theta(X)$ as the prediction. Then, supervised learning process aims to minimize the error between calculated result ($\hat{Y}$) and corresponding labels ($Y$) so that the model $H_\theta$ would be as accurate as enough to give Y with input of X. The cost function used in this study was mean-square-error (MSE) which was defined as $MSE = \frac{1}{m}\sum_i^m (\hat{Y} - Y)^2$. Here $m$ refers to the number of samples.

### Regression Neural Network Model

The regression neural network was developed as the second scheme for the detection purpose based on the architecture of multiple layer perceptron networks. Figure 5 showed the architecture of the network. The input layer was the ten selected features transformed from time domain of raw data by PCA. Then two hidden layers were defined with 10 and 4 neurons for each layer correspondingly and log-sigmoid function was used as the activation function for both layers. Output layer adopted the linear function as the activation function to return a single response to the output. The error function was same MSE function and gradient descent was used for optimization. Once the parameters for architecture was determined, hyper-parameters such as momentum term and adaptive learning ratio were used to optimize the gradient descent. Dataset 1 was divided into 3 subsets that 70% of data was used for training, 15% for testing, and 15% for the validation check. Dataset 2 and 3 were used for prediction only from the previously trained model. Lastly, stopping criterion for training was defined by reaching the thresholds of performance, gradient, or validation checks (figure 7).

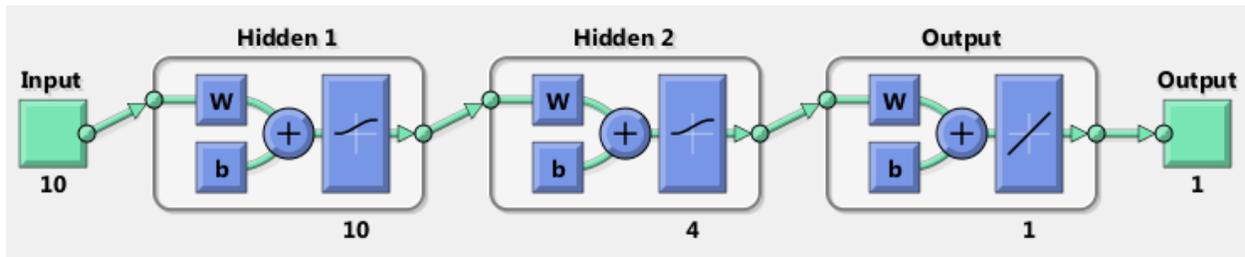

**Figure 5. Architecture developed for the multi-perceptron neural network**

## RESULTS

Figure 6 showed the results of the three datasets from four selected models. The training set (on the top row of figure 6) showed best match to labels comparing to other two datasets and the prediction was in the range of [0,1]. Figure 6 middle and bottom rows showed the performance for dataset 2 and 3 which we could observe that the general performance was acceptable in terms of location and ranges, but some missed predictions were also noticeable. Here, we also place the error table (Table 1) for the training dataset and the lower value indicated better performance of the model.

Figure 7 showed the performance of model training monitored across epochs. Figure 7 (a) showed the MSE for training, testing, and validation and lower MSE indicated better fitting of the model. The training was stopped at epoch 281 when it reached the validation check criterion. Figure 7 (b) showed the gradient, validation check, and learning rate across epochs. It was used to observe how the hyper-parameters performs during the training and perceives the potential local minima for gradient descent. Figure 7 (c) showed the error histogram for all samples (instance) in terms of training, testing, and validation subsamples. It was measured by subtracting labels (targets) from model predictions (outputs). Ideally, we would expect the error term closing to the zero-error line as much as possible.

Results shown in figure 8 illustrated the performance of trained model and predictions. Figure 8 (a) showed the prediction results for the training data set and we could find the predicted values ranged within the labeling range. But there were variations in space existed compared to the labels. Figure 8 (b) presented the prediction on dataset 2 by using the trained model. The appearance was close to the figure 6 (middle row) but with a clearer prediction on solid areas (blue region). Figure 8 (c) was the outcome for dataset 3 which also gave better detection on solid regions comparing to figure 6 (bottom row). All the predictions ranged for (b) and (c) remained close to the range of [0,1] which could be considered as a good sign for acceptable generalization of the model. Addition, predictions for dataset 1 and dataset 2 were failed to identify the region between two hollowed areas.

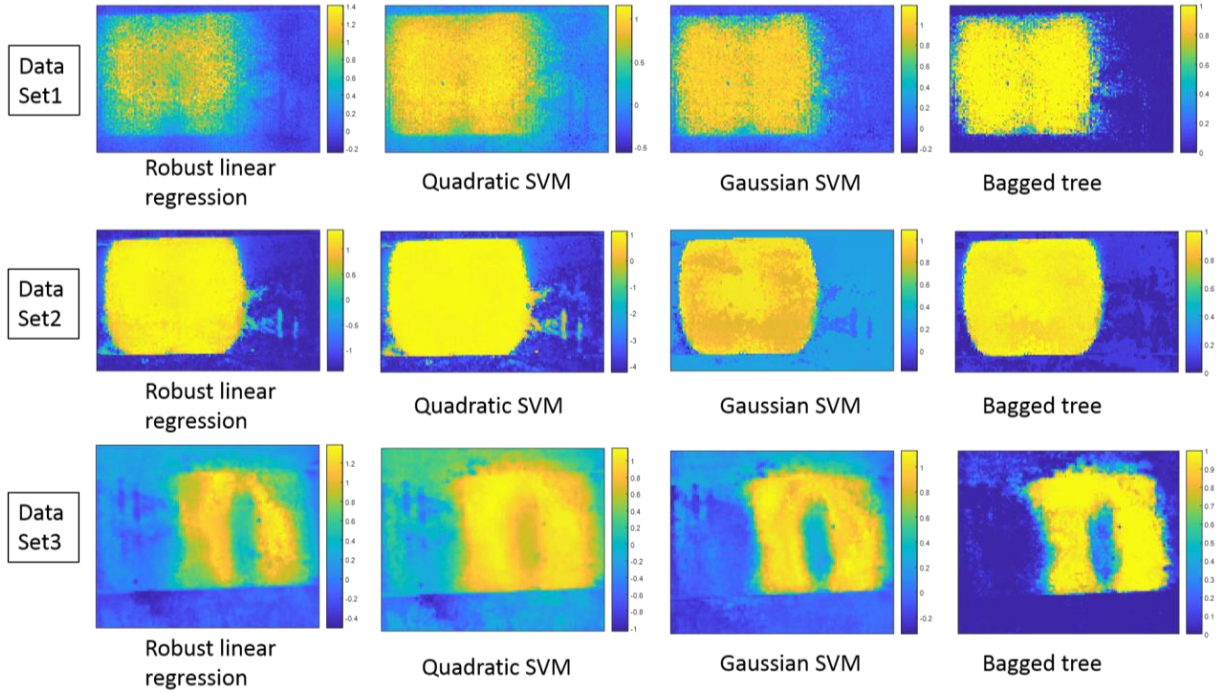

**Figure 6. Results from regression models for three datasets**

**Table 1. Training errors for selected models**

| Model Selected | Robust linear regression | Quadratic SVM | Gaussian SVM | Bagged tree |
|---|---|---|---|---|
| RMSE (Root mean square error) | 0.216 | 0.198 | 0.154 | 0.152 |

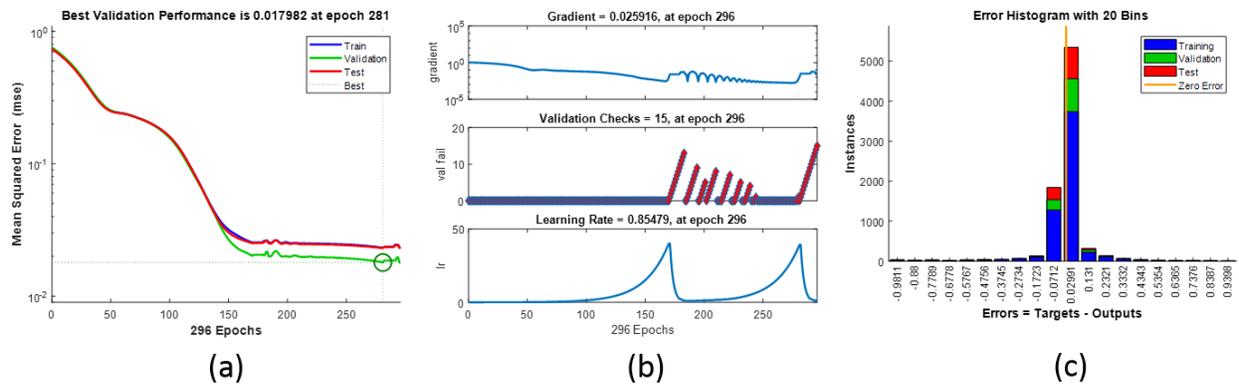

**Figure 7. Performance of model training**

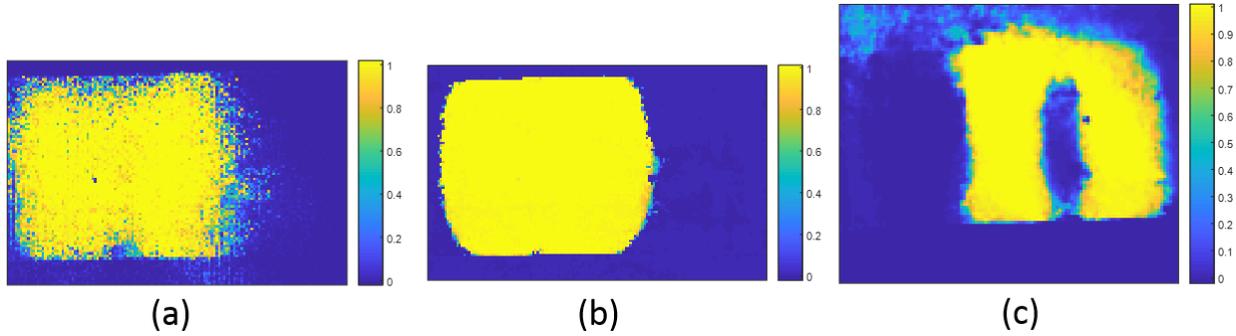

**Figure 8. Results for training and prediction from neural network**

## DISCUSSION AND CONCLUSION

According to the performance of the model prediction from two schemes, using time-series thermal feature could provide distinguish power for detecting the subsurface void-like defects under lab environment. There are several considerations need to be discussed: (1) this paper focused on the heating phase to mimic the reflection effect from a heating source. The temperature curve is different from cooling phase, but the feature is still distinguishable and will be conducted in the future study. (2) the feature transformed from the time domain by PCA can provide an additional solution to the problem when the data length is different. Like the three datasets used in this study, all of them differs in the length of frames (figure 4. (a)). Thus, the model trained for dataset 1 requires the same number of features as input and cannot directly be applied to dataset 2 and 3. However, after the PCA, all three datasets can use the same length of features (ten features used here). (3) normalizing the transformed feature before training turned out to be very important based on author's experimentation. (4) there is always a trade-off between the decreasing training errors and having a well-generalized model. Even though the author used three different datasets to illustrate the generalization of the model, the relatively small sample size still limits the establishment of good generalization of the model. However, the model could be improved as the training sample increases which is the advance of the machine learning itself. (5) comparing two schemes, neural network provides better performance for the regression learners.

In all, this paper presented a procedure to reconstruct the time-series thermal feature by using PCA and fit into supervised learning process to achieve pixel-wise detection for sub-surface void-like defects in a CMU block. Under a simple case setup with laboratory environment, two schemes (regression learner and neural network) were tested effective for detection purpose with appropriate tuning for paraments. Future studies will focus on implementing more sophisticated structures to handle more realistic cases under natural conditions.

## AUTHOR CONTRIBUTION STATEMENT

The authors confirm contribution to the paper as follows:

Study conception and design: C. Cheng, Z. Shen; data collection: C. Cheng; analysis and interpretation of results: C. Cheng, Z. Shen; draft manuscript preparation: C. Cheng. All authors reviewed the results and approved the final version of the manuscript.